\documentstyle[12pt]{article}

\begin{document}
\title{Large Scale Structure and Supersymmetric \\
Inflation Without Fine Tuning}

\author{ G. Dvali\\
Dipartimento di Fisica\\
Universita di Pisa and INFN\\
Sezione di Pisa, I-56100\\
Pisa, Italy\\
\and
Q. Shafi and R. Schaefer\\
Bartol Research Institute, University of Delaware\\
Newark, DE 19716, USA}

\date{ }
\maketitle

\begin{abstract}We explore constraints on the spectral index $n$ of density
fluctuations and the neutrino energy density fraction $\Omega_{HDM}$,
employing data from a variety of large scale observations.  The best fits
occur for $n\approx 1$ and $\Omega_{HDM} \approx 0.15 - 0.30$, over a range of
Hubble constants $40-60$ km s$^{-1}$ Mpc$^{-1}$.  We present a new class of
inflationary models based on realistic supersymmetric grand unified theories
which do not have the usual `fine tuning' problems.  The amplitude of
primordial density fluctuations, in particular, is found to be proportional to
$(M_X /M_P)^2$, where $M_X (M_P)$ denote the GUT (Planck) scale, which is
reminiscent of cosmic strings!  The spectral index $n = 0.98$, in excellent
agreement with the observations provided the dark matter is a mixture of
`cold' and `hot' components. \end{abstract}

\noindent PACS Nos: 98.80Cq, 98.65Dx, 12.10Dm, 11.30Pb

\newpage

Recent studies of large scale structure formation \cite{ss94}, when
confronted  with a variety of data from the Cosmic Background Explorer
\cite{wright94} (COBE) and other large scale galaxy surveys, provide support
for an inflationary scenario \cite{linde} in which the spectral index of
density fluctuations $n$ is close to unity and the dark matter is a mixture of
cold and hot components.  Non-supersymmetric grand unified theories which give
rise to precisely this scenario were constructed more than a decade ago
\cite{stecker}.  However, several fundamental challenges, including
unification of GUTS with gravity and the gauge hierarchy problem, strongly
hint that the supersymmetric grand unified (SUSY GUTS) framework may be a more
promising way to proceed.

Supersymmetric GUTS based on $G \equiv SU(3)_c \times SU(3)_L \times SU(3)_R$
have the desirable feature that they permit unification of the standard model
gauge couplings to occur at scales on the order of $10^{16} GeV$, which is
indicated by the recent LEP data.  Moreover, as has recently  been shown
\cite{dvali}, by introducing either some symmetries, or
alternatively R-symmetry, the  gauge hierarchy (`fine tuning') problem can be
resolved in SUSY GUTS based on $G$.  It  should be stressed that this
is accomplished without complicating the  `higgs sector' of the theory.  The
`minimal' choice will do, which is well  nigh impossible in GUTS such as
$SU(5)$ or $SO(10)$.  The additional symmetries also ensure that the
proton is essentially stable.

Encouraged by these developments, we investigate here if the inflationary
scenario can be realized within the framework of $G$.  We would call the
attempt
`successful' if the following conditions are met.  First, the scalar (higgs)
sector of the theory, including the inflaton part, is determined by particle
physics considerations.  Second, no `fine tuning' of parameters is needed.
[This includes the gauge hierarchy problem.]  Finally, the scenario must
be such that the Planck scale corrections from supergravity and/or
superstrings can be safely ignored.  Although non-trivial, it turns out that
all of these constraints can be satisfied within the framework of $G$.  The
inflationary scenario we are led to has previously been considered in general
terms by Linde \cite{linde93} (see also ref. \cite{liddle}), and been dubbed
`hybrid' inflation.  Our realization of `hybrid inflation' within a
supersymmetric
framework is unique in a number of ways and can be implemented in a variety of
models, especially those based on $G$.  One particularly important result has
to do with the amplitude of primordial density fluctuations, which turns out
to be
proportional to $(M_X / M_P)^2$, where $M_X$ denotes a superheavy (GUT scale)
and $M_P \simeq 1.2 \times 10^{19}$ GeV is the Planck mass.  The spectral
index of the density fluctuations is very close to unity as required by the
observations.

We begin with a definition of the cosmological parameters in the simplest
inflationary models.
\begin{enumerate}
\item The density $\rho$ is the critical density $\rho_c = 3 H_0^2/(8\pi G)$,
where $H_0 = 100 h$ km s$^{-1}$ Mpc$^{-1}$ is the present value of the Hubble
constant and $G$ is Newton's gravitational constant.

\item We take $h=0.5\pm 0.1$, which corresponds to the range of $h$ allowed
by the combination of observations and constraints on the age of the
universe in critical density models with a vanishing cosmological constant.

\item We take the baryon fraction to be the midpoint of the range allowed by
big bang nucleosynthesis \cite{bbn}, $\Omega_{baryon} h^2 = 0.0125\pm 0.0025$,
where $\Omega_{baryon}\equiv \rho_{baryon}
/\rho_c$.

\item The remaining mass density is in the dark matter which is composed
of some mixture of cold (CDM) and hot (HDM) components.  The
latter is assumed to be ``lightly" massive (few eV) relic neutrinos.  The
relative concentrations of the two components is left to be determined by the
data.  We include the possibility that the dark matter may be all CDM.

\item The power spectrum, which is the Fourier transform of the
two-point density autocorrelation function, has the primordial form
$P(k)\propto k^n$, where $k$ is the amplitude of the Fourier wavevector, and
$n$ denotes the spectral index.  The value $n=1$ corresponds to the
Harrison-Zeldovich spectrum.

\end{enumerate}

The above set of parameters define a family of inflationary models.  Their
ultimate test comes from comparison to data.  Here we will extend the
treatment previously described  in Ref. \cite{ss94} to include larger values
of the density fluctuation spectral index $n$ and different values of the
Hubble constant.  We will give a brief summary of the procedure used there and
present the new results in Figure 1.  The data we use is summarized next.

We begin with the data from COBE.  The amplitude is most conveniently
characterized in a relatively $n$ independent way by the  expected
hexadecupole moment of the temperature anisotropy  $\delta T_4 = 12.8\pm 2.3\
\mu$K, where the error includes effects such as radiometer noise and cosmic
variance \cite{wright94}.  (This treatment is consistent with the recent
analysis of the COBE two year data \cite{gorski}.)  This data set probes the
spectrum on the largest scales ($\sim 10^3$ $h^{-1}$ Mpc).   For smaller
scales, we use the data obtained from large galaxy surveys, particularly that
of the  Infrared Astronomy Satellite (IRAS)  catalogue of galaxies.  In ref.
\cite{feldman93} this catalogue  was used to obtain a direct estimate of the
power spectrum in the range $30\  h/{\rm Mpc}< 2\pi/k < 300\ h$/Mpc.  The
resulting power spectrum is in good agreement with estimates from other
surveys  based on optically selected galaxies.   The IRAS catalogue of
galaxies has also  been used in a different way to determine the large scale
bulk velocity field.  We use the values given by the potential flow algorithm
(``POTENT") collaboration (see ref. \cite{dekel94}).

    There are important constraints on the power spectrum which come from
structures which have evolved so that their description is inherently
non-linear.  If we make some basic assumptions about the formation of these
structures,  we can use this data to constrain the amplitudes of the
fluctuations in the linear theory.  We will use two such pieces of information.
First, in pre-COBE days the  normalization of power spectra was usually done by
specifying the mass fluctuation  on $8\ h^{-1}$ Mpc scales, and much effort has
been made to determine this  normalization from the data.  In ref
\cite{white93}, an attempt was made to estimate this quantity from the
abundance of clusters.   Estimates of the mass fluctuation on $8\ h^{-1}$ Mpc
from different galaxy surveys yield results that are consistent
\cite{iras8}.  We adopt the  synthesized value
$\delta M/M(8\ h^{-1}$ Mpc) $<0.8$ as a constraint on the amplitude of linear
perturbations.

    Second, the number of early quasars at high redshift requires a minimum
amplitude of mass fluctuation.  This issue has recently been extensively
studied \cite{quasar}.  In Ref. \cite{ss94} two of the present authors
estimated the error on this amplitude introduced by  theoretical modeling
uncertainties and found that the lower limit on the amplitude of mass
fluctuations required to make the early quasar population is  $\delta M/M(0.6\
h^{-1}\ { \rm Mpc}) \geq 1.1\pm 0.2$.  We will use this constraint here as
well \cite{klypin}.

   The data are then analyzed in the following way.  For a given value of $n$
and a hot dark matter fraction $\Omega_{HDM}$,   we calculate up to a
normalization factor, the hexadecupole moment of relic temperature
fluctuations, the power spectrum for the IRAS power spectrum wavenumbers, the
large scale velocities, the linear mass fluctuation on 8 $h^{-1}$ Mpc, and the
amplitude of the quasar sized mass fluctuations.   In addition, while most of
our list of constraints depend on the mass fluctuations, the variations in
galactic number can be somewhat different than mass fluctuations.  To compare
our mass fluctuation power spectrum with galactic number we have to also
introduce a ``density bias  factor" $b_I$ to allow for the possibility the IRAS
galaxies may not trace  the mass.  The fact that the depth coordinate in the
galaxy survey is doppler redshift space rather than physical length introduces
some effects which must be corrected for (see ref. \cite{kaiser87}).  For each
value of $\Omega_{HDM}$  and $n$ we first do a least squares fit of the
normalization and the bias  factor $b_I$.

    After fitting these parameters, we proceed to calculate the value of the
$\chi^2$ (goodness of fit) statistic from the data for each model.  For
the bulk velocities, we have combined the uncertainty due to ``cosmic variance"
in quadrature with the quoted analysis errors, because the velocities
correspond to a very limited sampling of a Gaussian velocity field.  After
finding the best fitting model, we record how the $\chi^2$ statistic changes
as we vary the parameters $n$ and $\Omega_{HDM}$ and draw confidence level
contours in the $n-\Omega_{HDM}$ plane \cite{dchisq}.  We repeat this entire
procedure for different values of the Hubble constant.

   Figure one dramatically shows how the data favor values of $n$ very close
to unity with the actual limits depending somewhat on the mix of dark matter
and the Hubble constant.  Overall we can say that with 99\% confidence, $0.80
<n [H_0/(50 {\rm km}\ {\rm s}^{-1}\ {\rm Mpc}^{-1})]^{1/2} <1.15$, independent
of the dark matter composition.  The value of $n=1$ works well over the range
of Hubble constants $40-60$ km s$^{-1}$ Mpc$^{-1}$. It appears that $n$ is
constrained to be quite close to unity by the data.

We now discuss how such density fluctuations can be realized in realistic
supersymmetric GUTS.  We are particularly interested in identifying models in
which there are no `fine tuning' (including gauge hierarchy) problems.  To set
things up, consider the following globally supersymmetric renormalizable
superpotential $W$:  \begin{equation} W = \kappa\  S\  \bar{\phi}\  \phi - \mu
^2 S \end{equation}
where $\phi (\bar{\phi})$ denote a conjugate pair of superfields transforming
as non-trivial representations of some gauge group, while $S$ is a gauge
singlet superfield.  This superpotential is `natural' in the strong sense
\cite{witten}.  It is of the most general form consistent with R-symmetry
under which $S \rightarrow e^{i \gamma} S$, $W \rightarrow e^{i \gamma} W$,
while the product $\bar{\phi} \phi$ is invariant.  Note that cubic terms in
$\phi$ and $\bar{\phi}$ can be forbidden by assuming, for example, the
transformations $\phi \rightarrow \bar{e}^{i\gamma} \phi, \bar{\phi}
\rightarrow e^{i \gamma} \bar{\phi}$.  In realistic models such terms may be
allowed without altering the main conclusions.

We point out that, at least in the global SUSY case, the R-symmetry is the
unique choice for  implementing the `false' vacuum inflationary scenario in a
natural way.  It is the only symmetry which  can eliminate all of the
undesirable self-couplings of the (inflaton) $S$, while allowing the linear
term in the superpotential.  With supersymmetry unbroken, the potential takes
the form (we represent  the scalar components with the same symbols as the
superfields if there is no  danger of confusion!): \begin{equation} V(S,\phi,
\bar{\phi})
= \kappa^2 \mid S \mid ^2 \left[ \mid \phi \mid ^2 + \mid \bar{\phi}\mid^2
\right] + \mid \kappa \phi \bar{\phi} - \mu^2\mid^2 + D-\rm{terms}
\end{equation} The
$D$-terms vanish along the (D-flat) direction $\mid\phi\mid = \mid
\bar{\phi}^*\mid$.  Consequently, the only supersymmetric minimum of the
potential is at \begin{equation} \begin{array} {rcl} <S> & = &0 \\  \\ M_X
\equiv \langle\mid\phi\mid\rangle = \langle\mid \bar{\phi}\mid \rangle & = &
\mu / \sqrt{\kappa} (\mu>0, \kappa>0)\end{array} \end{equation}  We will say
more about the scale $M_X$ shortly.

Consider now an early universe scenario with chaotic initial conditions. For
$\mid S\mid > \mid S_c\mid = \mu / \sqrt{\kappa}$, the effective  potential
$V$ is minimized by \linebreak $<\phi>=<\bar{\phi}>=0$.  That is,  for
$|S|>S_c$, the energy density is dominated by the `false' vacuum energy
density $\mu^4$, which can therefore lead to an exponentially expanding
(inflationary) universe.  The potential in (2) does not contain a term which
can drive $S$ to its minimum value.  This, however, is no longer the case when
the quantum corrections are taken into consideration.

With $|S|>S_c$, both $\phi$ and $\bar {\phi}$ vanish and there is a nonzero
$F_S-term(=\mu^2)$ which breaks supersymmetry, such that the one loop
corrections to the effective potential are non-vanishing, and given by
\cite{coleman}
\begin{equation} \Delta V(S) = \sum_i{(-1)^F
\over 64 \pi^2} M_i(S)^4 ln({M_i(S)^2 \over \Lambda^2}) \end{equation} where
the
summation is over all helicity states, $(-1)^F$ indicates that the bosons and
fermions make opposite sign contributions, and $\Lambda$ denotes a
renormalization mass.  The quantum corrections will help drive $S$ to its
minimum.

Note that for $S>S_c$ there is no mass splitting inside the gauge
supermultiplets or the
$S$-superfield (actually the masses of $S$-scalar and its fermionic
superpartner both vanish).  The non-vanishing contribution is from the mass
splitting within the $\phi$ , $\bar{\phi}$ superfields. The complex scalars in
$\phi, \bar{\phi}$ are split by the nonzero $F_S$-term into two pairs of real
scalar and pseudoscalar  components with mass squared $\kappa^2 S^2 \pm \kappa
\mu^2$, whereas the fermionic partners have mass $\kappa S$.  The one loop
corrected effective potential (along the inflationary trajectory $S > S_c$,
$\phi =\bar{\phi} =0$) is given by  \begin{eqnarray} V_{eff}(S) = \mu^4 +
{\kappa^2 \over 32 \pi^2} [2 \mu^4 ln ({\kappa^2\mid S \mid^2 \over
\Lambda^2}) + \nonumber \\ (\kappa S^2 - \mu^2)^2 ln(1-{\mu^2 \over \kappa
S^2}) + (\kappa S^2 + \mu^2)^2 ln(1 + {\mu^2 \over \kappa S^2})] \end{eqnarray}
If $S$ is sufficiently greater than $S_c$, $V_{eff}(S)$ reduces to the simpler
form
\begin{equation}
V_{eff}(S\gg S_c) \approx \mu^4\left[ 1 +{\kappa^2 \over 32 \pi^2}({\rm
ln}{\kappa^2
S^2\over \Lambda^2} + {3\over 2})\right]
\end{equation}

For $|S|>S_c$, the inflationary phase is dominated by the false vacuum energy
$\mu^4$ as in the tree level case, but the additional contribution in (5) will
now drive $S$ to its minimum.  The GUT phase transition takes place only after
the $S$ field drops to its critical value $S_c(=M_X)$.  Below $S_c$, the $S$
field is driven to zero by the positive mass term $\kappa^2 |S|^2 |\phi^2|$
which is increasingly more effective due to the increase of the $\phi,
\bar{\phi}$ vevs (induced by the decreasing $S$).  All of the fields rapidly
adjust to their vacuum values (3), thereby restoring supersymmetry.

Note that the end of inflation does not necessarily coincide with the GUT
phase transition which occurs when $S$ approaches $S_c$. The end is signaled
when the ``slow roll" condition is violated for some $S>S_c$. We can
characterize the ``slow roll" condition as (see first paper in ref.
\cite{liddle})
\begin{equation} \epsilon << 1, |\eta | << 1
\end{equation} where  \begin{equation} \epsilon = {M_P^2 \over 16 \pi }({
V^{\prime} \over V})^2,
\eta = { M_P^2 \over 8 \pi} {V^{\prime \prime} \over V} \end{equation} (the
prime refers
to derivatives with respect to $S$).  The inflationary phase may end before
the GUT transition if the above conditions are violated at some $S>S_c$. For
convenience, we can use the parametrization $S=xS_c$, where the parameter $x$
characterizes the rolling of $S$. (The GUT phase transition occurs for $x=1$.)
The quantities $\epsilon$ and $\eta$ are given by
\begin{eqnarray}
\epsilon = ({\kappa^2 M_P \over 16 \pi^2 M_X})^2 {x^2 \over 16 \pi}
[(x^2 - 1)ln(1-{1 \over x^2}) + (x^2 + 1)ln(1 + {1 \over x^2})]^2 \nonumber\\
\eta = ({ \kappa M_P \over 4 \pi M_X})^2 {1 \over 8 \pi}
[(3x^2-1)ln(1- {1 \over x^2}) + (3x^2 +1)ln(1 + {1 \over x^2})]
\end{eqnarray}
Note that $\eta$ becomes infinitely large for $x=1$, so that inflation ends as
$x$ approaches 1 (from above).

So far, we have not introduced any supersymmetry violation in the system (the
global minimum in (3) is supersymmetric).  In conventional schemes (say N=1
supergravity), this breaking is introduced through the soft SUSY violating
terms in the tree level potential. The main influence of such terms on the
inflationary scenario discussed above arises from the fact that the SUSY
breaking induces a TeV scale  $({\rm mass})^2$-term for the scalars, in
particular for the $S$ field. The term $m^2|S|^2$ ($m\sim $ TeV) provides an
extra
force driving $S$ to the minimum. However, unless the coupling constant
$\kappa$ is very small, the soft mass terms only provide a small correction to
$V_{eff}(S)$ in (5), and so cannot significantly affect the above dynamics.
This is
not surprising since for $|S|>S_c$, the non-supersymmetric $(mass)^2$ splitting
inside the $\phi ,\bar{\phi}$ superfields is $\kappa \mu^2$ which, as we shall
see, is much larger then $m^2$. In such a situation the inflationary scenario
above is practically independent of the particular mechanism of supersymmetry
breaking.

Let us now compare the predicted quadrupole anisotropy, based on (5),  with
the value ($\approx 7\times 10^{-6}$) measured by COBE.  From the scalar
density fluctuations one has (see first
paper in ref. \cite{liddle})
\begin{equation}\begin{array}{rcl} \left( {\Delta T\over T}\right)_Q & \approx
& \sqrt{32 \pi \over 45} {V^{3 \over 2} \over V^\prime M_P^3} \mid_{x{_Q}}\\ &
\approx & (8 \pi N_Q)^{1 \over 2}  (M_X / M_P )^2 \end{array} \end{equation}
where the subscript $x_Q$ indicates the value of $S$ as the scale (which
evolved to the present horizon size) crossed outside the de Sitter horizon
during inflation, and $N_Q (\approx 50 -60)$ denotes the appropriate number of
e-foldings.  The formula in (10) is remarkable in that the fluctuation
amplitude is proportional to $(M_X/M_P)^2$, just as in the cosmic string
scenario!  The amplitude turns out to be in  the right ball park, without
having to `fine tune' additional parameters (such  as dimensionless quartic
couplings and/or the mass of the inflaton).  Using  (10), we can estimate the
fundamental parameter $M_X$ to be on the order of  $10^{15.5} GeV$. We have
ignored the contribution of the tensor fluctuations to the anisotropy, because
they are suppressed by a factor of $\kappa/[8\pi (N_Q)^{1/2}]$ relative to
the scalar component in eq. (10).

The spectral index $n$ of the density fluctuations is given by
\begin{equation} n \simeq 1 - {1 \over N_Q} \simeq 0.98 \end{equation}
which, as  we have seen earlier, is in the central range of the values
preferred by observations.

An estimate of the coupling $\kappa$ is obtained from the relation
\begin{equation} {\kappa \over x_Q} \sim {8 \pi^{3/2} \over \sqrt{N_Q}}
{M_X \over M_P} \end{equation}
With $x_Q \sim 10$ say, (which corresponds to $S\sim 10^{16.5}$ GeV, and the
Planck scale corrections can be safely ignored), the
coupling $\kappa$ turns out to be on the order of $10^{-2}$.
Note that for this value of $\kappa$, the tensor generated anisotropies are
less than $10^{-4}$ of the scalar anisotropy amplitude.

Having outlined how supersymmetric models can lead to a successful
inflationary scenario without invoking small dimensionless couplings, we now
proceed to discuss how this idea can be realized in realistic SUSY GUTS.  We
would like the model to be well motivated from the particle physics viewpoint.
For instance, a non-trivial constraint would be the absence of the gauge
hierarchy and proton decay problems.  Although this constraint presumably can
be met in GUTS such as $SU(5)$ or $SO(10)$, the higgs sector of the theory
becomes quite messy.  An even more formidable constraint stems from the fact
that
the phase transition involving the gauge non-singlet fields $(\phi,
\bar{\phi})$  occurs at the end of inflation.  In $SU(5)$ this would lead to
the monopole problem.  We therefore must resort to SUSY GUTS with rank 5 or
higher in order that at least the possibility exists for the monopoles to
be inflated away.

  The simplest  example of SUSY GUTS we are aware of in which the gauge
hierarchy problem  can be resolved with a minimal higgs system is based on $G
(\equiv SU(3)_c  \times SU(3)_L \times SU(3)_R$).  Under the gauge group $G$,
the
left handed lepton, quark and antiquark  superfields respectively transform as
$(1, \bar{3}, 3), (3,3,1)$ and $(\bar{3}, 1, \bar{3})$.  They are denoted as
$\lambda_i, Q_i$ and $Q^c_i (i = 1,2,3)$: \begin{equation} \lambda_i = \left(
\begin{array} {ccc} H^{(1)}&H^{(2)}&L\\e^c&\nu^c&N\end{array}\right)$$ $$Q_i =
\left( \begin{array} {c} u\\d\\g \end{array}\right)\end{equation} $$Q^c_i =
(u^c \ d^c \
g^c)$$ Here $H^{(1)}, H^{(2)}$ and $L$ denote $SU(2)_L$ doublet superfields,
$e^c$ is an $SU(2)_L$ singlet, $N$ and $\nu^c$ are standard model singlets,
while $g(g^c)$ denote additional down-type quark (antiquark) superfields.  The
$SU(3)_L$ ($SU(3)_R$) symmetry acts along columns (rows). The symmetry
breaking of $G$ to $SU(3)_c \times U(1)_{em}$ requires at least two sets of
higgs supermultiplets, $\lambda + \bar{\lambda}$ and $\lambda^{\prime} +
\bar{\lambda}^{\prime}$, where $\lambda(\lambda^{\prime})$ transform as
$\lambda_i$ above.  The conjugate fields
$\bar{\lambda}(\bar{\lambda}^{\prime})$ are needed to ensure that the SUSY
breaking scale is well below the GUT scale and that the anomalies cancel.

Imposition of a suitable R-symmetry, largely determined by
the requirement that a pair of electroweak doublets (in $\lambda$) remains
light, fixes the renormalizable part of the superpotential as
follows (we will not give the couplings involving the chiral
fermion families):

 \begin{eqnarray}                              W & =  & W_{\lambda} +
W_{\lambda^{\prime}} +          W_{\lambda \lambda^{\prime}}  \nonumber
\\                    W_{\lambda} & =  & \kappa S \lambda \bar{\lambda}
-                       \mu^2 S+ a \bar{\lambda}^3 \nonumber\\
W_{\lambda^{\prime}} & = & b \lambda^{\prime 3} +            c
\bar{\lambda}^{\prime 3} \nonumber \\   W_{\lambda \lambda^{\prime}} & = & d
\bar{\lambda}^2 \bar{\lambda}^{\prime}         + e \bar{\lambda}
\bar{\lambda}^{\prime 2} \end{eqnarray} Here the bilinear and trilinear terms
stand for the appropriate $G$ invariant combinations.  For instance, $\lambda
\bar{\lambda} \equiv Tr (\lambda^A_{\alpha} \bar{\lambda}^{\alpha}_A) ,
\lambda^{\prime 3} \equiv \epsilon ^{\alpha \beta \gamma} \epsilon _{ABC}
\lambda^{\prime A}_{\alpha} \lambda^{\prime B}_{\beta}
\lambda^{\prime C}_{\gamma}$, etc., where the
Greek (Latin) indices refer to $SU(3)_L (SU(3)_R)$.  The R-charges of the
superfields are as follows:  $S(1)$, $\lambda(-{1 \over 3})$,
$\bar{\lambda}({1 \over 3})$, $\lambda^{\prime}({1 \over 3})$,
$\bar{\lambda}^{\prime}({1\over 3})$, where $W\rightarrow e^{i\theta}W$ under
R. The superlarge vevs will be oriented
along the directions \begin{equation} \mid \lambda \mid = \mid \bar{\lambda}^*
\mid = \left( \begin{array} {ccc}
0&0&0\\0&0&0\\0&0&N\end{array}\right)\end{equation} \begin{equation} \mid
\lambda^{\prime} \mid = \mid \bar{\lambda}^{\prime *} \mid = \left(
\begin{array} {ccc} 0&0&0\\0&0&0\\0&\nu^{c
\prime}&0\end{array}\right)\end{equation} The absence of a $\lambda^3$ term
in (14) guarantees that the electroweak doublet pair $(H^{(1)} - H^{(2)})$in
$\lambda$ remains massless at tree level, despite the GUT symmetry breaking.
The well known `$\mu$-term' for this pair is generated by the
non-renormalizable coupling $\lambda^3(\lambda^\prime \bar{\lambda^\prime})^3
/ M^6 , (M \equiv M_P / \sqrt{8 \pi} \sim 10^{18} GeV)$ which is the leading
contribution allowed by the R-symmetry.  With $<\lambda^{\prime}>/M \sim
10^{-2}$, $\mu$ is in the TeV range as desired.   [Note that without the
$W_{\lambda, \lambda^{\prime}}$ term, the superpotential in (14) would be
invariant
under $[SU(3)_L \times SU(3)_R]^2$.]

It is important to point out that in addition to resolving the gauge hierarchy
problem, the R-symmetry guarantees the absence of dimension four and five
baryon number violating operators.  That is, cubic superpotential couplings
such as $Q_i Q_j Q_k$ are absent, where the $Q_i$'s and $Q_i^c$'s carry
R-charges $= {2 \over 3}$.  The proton is essentially stable according to this
model. Moreover, the desired couplings such as $Q_i Q^c_j \lambda$, which
provide masses for the quarks, are allowed.   Similarly, there are no
undesirable lepton number violating couplings.

The superpotential (14) is the generalization of (1), on which the
inflationary scenario was based,  to a realisitic SUSY GUT model. Without
going into the details, which will be published elsewhere, we briefly summarize
how things work out.  The role of the $\phi - \bar{\phi} - S$ sector in (1) is
played by the superfields $\lambda - \bar{\lambda} - S$.  [Note that the
$\bar{\lambda}^3$ term in (14) vanishes along the vacuum direction (15).]
Prior to inflation, the vevs in the
$\lambda^{\prime} - \bar{\lambda}^{\prime}$ sector, which arise
from an interplay of suitable non-renormalizable couplings and supersymmetry
breaking mass terms,
break the $(SU(3))^3$ GUT symmetry to $SU(3)_c \times SU(2)_L \times
SU(2)^{\prime}_R \times U(1)$.  The subsequent inflationary phase driven by
the $S$ field inflates away the monopoles produced during the first
$(\lambda^{\prime} - \bar{\lambda}^{\prime})$ breaking.  The $\lambda -
\bar{\lambda}$ fields acquire non-zero vevs $(\sim 10^{15.5}$ GeV) at the end
of inflation.

  An inflationary scenario is incomplete without some discussion of i) how the
baryon asymmetry is generated and ii) the dark matter candidate(s).  As far as
(i) is concerned, it has been shown in ref. \cite{dvali} that the resolution of
the
gauge hierarchy problem within the framework of $G$ necessarily leads to an
additional pair of electroweak doublets.  (Actually, one effectively ends up
with a $5 + \bar{5}$ of $SU(5)$, of mass $\sim TeV$.  Unification of the three
gauge couplings at scales $\sim 10^{16} GeV$ is thereby preserved.) This
strongly suggests generating the baryon asymmetry at the electroweak scale.
The reheat temperature in this scheme is high, $\sim 10^{14}$ GeV, which
means that the gravitino mass must be large in order that the
primordial gravitino problem is avoided.  Precisely how this works out would
depend on how local supersymmetry is broken.

As far as dark matter is concerned, the lightest supersymmetric particle (LSP)
is completely stable and so presents itself as an ideal `cold' dark matter
candidate.  Indeed, one can speculate here about the nature of the LSP.  The
point is that in the absence of the two lighter families, the well known
parameter $tan\beta \ (\equiv <H^u>/<H^d>)$ of the MSSM is equal to $m_t/m_b$
at the GUT scale $M_X$.  (This follows from the superpotential coupling $Q\
Q^c\ \lambda$.)  Coupled with the radiative electroweak breaking scenario,
this leads to a cold dark matter candidate essentially composed of the bino
with mass in the range of 200-350 GeV.

What about the neutrinos in this scheme?  The lepton chiral supermultiplets
$\lambda_i (i = 1,2,3)$ transform according to the same $G$ representation as
the higgs supermultiplets $\lambda, \lambda^{\prime}$.  The R-symmetry allows
the coupling $\lambda_i \lambda_j \lambda$ provided that the R-charge of
$\lambda_i$ is ${2 \over 3}$.  This is needed to provide mass say to the
$\tau$ lepton.  But it also gives a Dirac mass $m_{\tau} tan\beta$ to the
$(\nu_{\tau}$ and $\nu^c_{\tau})$ neutrinos in $\lambda_i$.  Now, the
R-symmetry prevents one from giving a large Majorana mass to the $\nu^c$
fields, which could make the neutrinos unacceptably heavy.  We are therefore
led to introduce three G-singlet superfields $F_i$ (one per chiral family) in
order that the standard see-saw mechanism can be implemented.  One of the
neutrinos can now have a mass in the few electron volt as suggested by the
data on large scale structure formation.

  In conclusion, it is gratifying to learn that a satisfactory inflationary
scenario, consistent with large scale structure observations, can be realized
in realistic supersymmetric GUTS in which the sector that `drives' inflation is
determined by particle physics considerations, and no fine tuning of the
parameters is needed.
\medskip

{\bf Acknowledgement.}   We would like to gratefully acknowledge the support
of this research by the DOE under grant DEFG02 - 91ER40626.

\section{Figure Captions}

Figure 1.  The $\chi^2$ contours (68 \%, 95\%, and 99\% confidence levels)
for values of $n$ and $\Omega_{hdm}$ implied by the large scale structure
data.  In the left panel we have the constraints for a Hubble constant of $40$
km s$^{-1}$ Mpc$^{-1}$.  Similarly we have the constraints for Hubble constants
of $50$ and $60$ km s$^{-1}$ Mpc$^{-1}$ in the center and right panels.
Values of $n\sim1$ are consistent with all values of the Hubble constant.  We
note that decreasing the Hubble constant favors larger values of $n$ and
smaller $\Omega_{HDM}$.

\end{document}